# A Sustainable and Reward Incentivized High-Performance Cluster Computing for Artificial Intelligence: A Novel Bayesian-Time-Decay Trust Mechanism in Blockchain[1]

**Abstract:**

In an age where sustainability is of paramount importance, the significance of both high-performance computing and intelligent algorithms cannot be understated. Yet, these domains often demand hefty computational power, translating to substantial energy usage and potentially sidelining less robust computing systems. It's evident that we need an approach that is more encompassing, scalable, and eco-friendly for intelligent algorithm development and implementation.

The strategy we present in this paper offers a compelling answer to these issues. We unveil a fresh framework that seamlessly melds high-performance cluster computing with intelligent algorithms, all within a blockchain infrastructure. This promotes both efficiency and a broad-based participation. At its core, our design integrates an evolved proof-of-work consensus process, which links computational efforts directly to rewards for producing blocks. This ensures both optimal resource use and participation from a wide spectrum of computational capacities.

Additionally, our approach incorporates a dynamic 'trust rating' that evolves based on a track record of accurate block validations. This rating determines the likelihood of a node being chosen for block generation, creating a merit-based system that recognizes and rewards genuine and precise contributions. To level the playing field further, we suggest a statistical 'draw' system, allowing even less powerful nodes a chance to be part of the block creation process.

In this age of rapid technological advancement, it's essential to strike a balance between leveraging powerful computation for intricate AI challenges and promoting widespread participation. By doing so, we aim to cultivate a more equitable and sustainable AI development landscape. Our aspiration is to pave the way for an environmentally conscious and inclusive AI progression, minimizing its ecological footprint while maximizing the benefits shared across the board.

Our proposed strategy melds the capabilities of high-performance cluster computing with the nuances of artificial intelligence. Central to this is an inventive reward structure closely aligned with the computational might one brings to the table. Drawing inspiration from the blockchain realm, we've harnessed the Proof-of-Work consensus mechanism, creating a direct correlation between the computational effort and the rewards from block production. The ensuing sections will unpack this idea further, diving deep into validation processes, trust-building measures, and strategies to motivate even those with limited computational resources.

---

[1] Corresponding Author: M. Murat Yaşlıoğlu, Istanbul University, School of Business, murat@yaslioglu.com

**Keywords**

High Performance Cluster Computing, Artificial Intelligence, Blockchain, Consensus Mechanism, Time-Decay Trust, Bayesian Approach, Sustainability

**Introduction**

High Performance Cluster Computing (HPCC) refers to the use of clusters of computers to perform high-performance computing tasks. A cluster is a group of interconnected computers that work together to solve complex computational problems. HPCC systems are designed to provide high computational power, storage capacity, and network bandwidth to handle large-scale and computationally intensive applications (Foster et al., n.d.)

The use of clusters in HPCC allows for parallel processing, where multiple tasks are divided among the cluster nodes and executed simultaneously, resulting in faster computation times. This parallelization is particularly beneficial for scientific and engineering workloads that require intensive calculations and simulations (Kindratenko et al., 2009).

The performance and scalability of HPCC systems are dependent on the underlying hardware and software components. High-performance computing clusters often incorporate specialized hardware, such as Graphics Processing Units (GPUs), to accelerate computations (Kindratenko et al., 2009). Additionally, efficient task scheduling algorithms are crucial for maximizing the utilization of cluster resources and achieving high performance (Singh & Singh, 2015).

In summary, High Performance Cluster Computing (HPCC) involves the use of clusters of interconnected computers to perform computationally intensive tasks. It finds applications in various sectors, including electric vehicle scheduling and media convergence. HPCC systems leverage parallel processing and specialized hardware to achieve high computational power and efficient resource utilization. Effective task scheduling algorithms are essential for optimizing performance in HPCC environments (Foster et al., n.d.; Kindratenko et al., 2009; Naylor et al., 2019; Singh & Singh, 2015).

High Performance Cluster Computing (HPCC) has undeniably been a cornerstone in pushing the boundaries of computational abilities, as cited by Dongarra et al. in 2003. Yet, as we delve deeper into the age of AI and intricate deep learning designs, there's a marked surge in computational necessities, as highlighted by Parmar et al. in 2023. Traditional computing setups often find themselves outpaced by the intricate demands of advanced AI paradigms, which lean heavily on parallel processing faculties to manage vast data and intricate operations.

Recognizing this gap, innovators have introduced bespoke hardware accelerators tailored to support swift computations for data-rich tasks, as discussed by Parmar et al. in 2023. Such equipment champions parallel matrix calculations and adeptly holds vast quantities of data, as observed by Sahu et al. in 2022. One fascinating specimen in this realm is the optoelectronic artificial synapse. It gracefully marries optical with electrical impulses to finesse device conductance, thereby enriching synaptic operations. Sahu et al. in 2022 note the perks it brings to the table: heightened bandwidth, exceptional propagation speed, and resistance to crosstalk, effectively side-stepping the pitfalls of traditional neuromorphic circuits.

Furthermore, the age-old Von-Neumann computing blueprint, characterized by its discrete memory and processing chambers, grapples with quandaries in the face of cutting-edge AI designs. Issues like the Von-Neumann bottleneck and the looming memory wall can stifle the adept execution of these algorithms, as pointed out by Sahu et al. in 2022. Offering a glimmer of hope, neuromorphic computing, which takes cues from the neural constructs of our brain, emerges as a commendable alternative. Its prowess in mirroring intricate learning activities while being power-thrifty could herald a new era in AI-informed tech, as Sahu et al. elucidate in 2022.

Venturing into the realm of environmental stewardship, it's evident that AI and Machine Learning (ML) approaches have reshaped the strategies employed to guard Mother Nature, as per Wearn et al. in 2019. From forecasting a species' risk of fading into oblivion, gauging the global sway of fisheries, to sifting through wildlife telemetry data, ML algorithms are proving invaluable. Lending further momentum to this movement, endeavors like Microsoft's 'AI for Earth' and Google's 'AI for Social Good' have come forward with supplemental assets and methodologies to address nature conservation with a touch of AI intelligence, as underscored by Wearn et al. in 2019.

In essence, the soaring computational needs of AI and deep learning models underscore the gaps in conventional computing setups (Parmar et al., 2023). Emerging technologies, like optoelectronic artificial synapses, step in to bridge this gap, championing parallel matrix calculations and adept data storage capabilities (Sahu et al., 2022). On another front, neuromorphic computing, inspired by our brain's neural construct, provides a beacon of hope against the constraints of the conventional Von-Neumann computing mold (Sahu et al., 2022). Turning our gaze towards environmental stewardship, it's evident that AI and ML methodologies have been harnessed to tackle diverse ecological challenges, receiving a boost from significant corporate initiatives (Wearn et al., 2019). In 2022, Arief Ginanjar and Kusmaya Kusmaya pinpointed the dependability of HPCC, especially in settings that employ programming tongues with virtual machine ecosystems like Java.

To wrap things up, weaving HPCC into distributed ecosystems has been a game-changer in managing colossal AI missions. This includes nurturing intricate neural layouts and navigating torrents of real-time data. It's the scalability and prowess of HPCC that has stood front and center, making these ambitious endeavors feasible (Das et al., 2017).

**Literature Review**

The allure of blockchain technology, praised for its decentralized essence and unwavering character, resonates across diverse sectors. A key highlight within its arsenal is the consensus mechanism, a linchpin that upholds data veracity and reliability. A deep dive by Shivani Wadhwa and colleagues in 2022 illuminated the nuances of varied blockchain consensus methods like proof of work (PoW) and proof of stake (PoS). Their discourse further introduced an eco-friendly consensus strategy, amplifying the energy stewardship within blockchain frameworks.

Leveraging the consensus mechanism of blockchain as a reward system can motivate network members, spurring vibrant involvement while cultivating mutual trust and synergy (Patil, 2023). Such an incentivized structure finds its stride especially in HPCC terrains dedicated to AI. Here, nodes or collaborators stand to gain rewards tied to their computational endeavors, promising judicious resource distribution and nimble data management (Patil, 2023).

Indeed, blockchain's influence extends to arenas like the Internet of Things (IoT), reputation paradigms, and the financial cosmos (Patil, 2023). It carves its niche as an unalterable ledger championing decentralized exchanges (Patil, 2023). Yet, like any pioneering technology, blockchain grapples with its set of teething troubles, including scalability conundrums and security concerns (Patil, 2023). For a panoramic view of the blockchain realm, spanning its consensus blueprints to architectural insights, one can delve into the scholarly piece by (Patil, 2023).

Pivoting to the world of mobile ad-hoc networks (MANETs), recent innovations suggest roping in blockchain-anchored trust management systems to mitigate security quandaries (Lwin et al., 2020). Here, blockchain steps in as a fortified, credible podium to tackle security intricacies intrinsic to MANETs. Embedding blockchain within MANETs paves the way for inviolable trust frameworks guiding routing nodes (Lwin et al., 2020). This fusion also unlocks the potential for transparent data exchanges among autonomous entities. Empirical evidence underscores the efficacy of such blockchain-infused trust management in MANETs, highlighting slashes in validation durations and operational overheads, thereby turbocharging the network's holistic efficacy (Lwin et al., 2020).

In the realm of wireless sensor networks (WSNs), blockchain technology has been integrated to bolster trust management and enhance anomaly detection. Leveraging blockchain within WSNs cultivates trust among dispersed nodes, leading to a decentralized trust-oriented system. This unique attribute of blockchain optimally trims the expenses associated with sculpting and upkeeping trust for individual nodes in WSNs (Yang et al., 2021). Moving into the vehicular network landscape, blockchain-grounded decentralized trust mechanisms have been advanced, given the essential trust prerequisites of these networks, courtesy of their innate high mobility and dynamic characteristics (Lwin et al., 2020).

Shifting the spotlight to the supply chain territory, blockchain emerges as a contender to augment trust dynamics and traceability of information. Nonetheless, the sector yearns for more intricate studies to decipher blockchain's influence, especially concerning trust dynamics within supply chain matrices (Chen & Su, 2022).

In sum, the incorporation of blockchain both as a reward mechanism and for trust management across diverse networks – spanning HPCC for AI, MANETs, WSNs, and supply chains – has demonstrated encouraging outcomes, particularly in boosting engagement, fortifying security, and nurturing trust bonds (Patil, 2023; Lwin et al., 2020; Yang et al., 2021; Chen & Su, 2022). Despite the strides, hurdles like scalability and security loom large, demanding deeper dives and technological advancements in blockchain to truly harness its multifaceted potential across network domains.

Blockchain's inherent decentralized blueprint guarantees data distribution across numerous nodes, bolstering its resilience against manipulations and digital threats (Patil, 2023). Such robustness is invaluable in AI terrains, where data sanctity and defense mechanisms reign supreme. Further, the clear and auditable trails of blockchain transactions instill an extra layer of trust, assuring that the results emanating from AI models stand scrutiny (Salah et al., 2019).

Blockchain's gravitational pull is felt across arenas like the Internet of Things (IoT), reputation architectures, and fiscal services, where it etches its mark as an unyielding ledger championing decentralized interactions. Yet, the challenges of scalability and security remain persistent adversaries. A holistic perspective on blockchain's universe, unfolding its structural intricacies and consensus frameworks, is encapsulated in the scholarly work by (Patil, 2023).

Merging AI with blockchain yields what's termed as decentralized AI, empowering processing and analytical feats on authenticated and safeguarded data ensconced within the blockchain (Salah et al., 2019). This paradigm sidesteps the necessity for third-party arbiters or go-betweens (Salah et al., 2019). Consequently, blockchain is esteemed as a reliable repository for the colossal data reservoirs that AI frameworks often necessitate (Salah et al., 2019).

To encapsulate, the decentralized essence of blockchain, combined with its impregnable nature and open transaction records, earmarks it as the gold standard for underpinning data veracity and fortifications within AI initiatives (Patil, 2023; Salah et al., 2019). Yet, impending challenges, notably around scalability and security (Patil, 2023), call for dedicated research and innovation in blockchain to realize its optimal potential in AI's expansive landscape.

**Consensus Mechanisms in Blockchain**

Blockchain technology, often hailed as a revolutionary force in the world of decentralized systems, has garnered significant attention for its potential to transform various sectors. At the heart of this technology lies the consensus mechanism, a critical component that ensures the integrity, security, and reliability of data stored on the blockchain. This article delves into the recent advancements and research surrounding consensus mechanisms in blockchain.

In a decentralized system like blockchain, where multiple participants (or nodes) maintain the ledger, it's crucial to have a system in place to agree on the validity of transactions. This agreement system is termed as the consensus mechanism. It ensures that all participants in the network have a consistent version of the ledger, thereby preventing double-spending, fraud, and ensuring data integrity.

Recent Advancements and Research

1. Electric Vehicle Scheduling with Blockchain: With the rising demand for electric vehicles, there's a pressing need for efficient scheduling mechanisms at charging stations. Kakkar et al. (2022) proposed a novel blockchain and IoT-based consensus mechanism named COME for secure and trustable electric vehicle scheduling at charging stations. This mechanism aims to resolve conflicts at charging stations, ensuring a seamless experience for both the vehicle owner and the charging station.

2. Primary Node Election in PBFT: The Practical Byzantine Fault Tolerance (PBFT) consensus mechanism is renowned for its resilience against malicious nodes. Xie et al. (2022) introduced a primary node election method based on the probabilistic linguistic term set with a confidence interval (PLTS-CI) to enhance the efficiency of reaching consensus in PBFT. This method offers an accurate solution for primary node election by considering complex voting attitudes.

3. Convergence Media with Blockchain: The media industry has witnessed significant evolution, from traditional media to convergence media. Hu and Wang (2022) explored the use of blockchain for convergence media and introduced a consensus mechanism named proof of efficiency (PoE). This mechanism stimulates node activity and addresses the generation of the Matthew effect, ensuring a decentralized and efficient media ecosystem.

4. Product Traceability with Blockchain: Blockchain's traceability feature is particularly beneficial for product tracking. Kang et al. (2022) conducted research on blockchain product traceability and introduced a consensus mechanism for trusted data analysis. Their findings emphasize the benefits of blockchain in ensuring product authenticity and traceability.

At the heart of blockchain networks lie consensus mechanisms, integral to both their operation and security (Kokoris-Kogias et al., 2018). With the tech landscape in constant flux, the academic community is delving into cutting-edge consensus strategies tailored to sector-specific hurdles (Afzaal et al., 2022). The horizon of blockchain's impact, fortified by these mechanisms, spans realms as diverse as electric vehicle coordination to the fusion of media channels (Islam et al., 2023).

Shifting focus to crowdsourcing paradigms, consensus algorithms rooted in blockchain have emerged as solutions to the intertwined concerns of trust, security, and accountability. These blueprints are crafted to tackle predominant challenges, be it credible data sourcing, passive participation, or misinformation dissemination. Capitalizing on blockchain's decentralized ethos and its transparent framework, these consensus designs forge a resilient and trust-infused scaffold for crowdsourcing ventures (Afzaal et al., 2022).

Blockchain's influence, augmented by consensus methodologies, permeates a myriad of sectors. Embodying attributes like decentralization, autonomy, incorruptibility, and transparency, blockchain stands as a beacon of transactional security and precision (Islam et al., 2023). The potential of such blockchain-infused applications is boundless, heralding innovations in sectors like green energy platforms (Nepal et al., 2022), digital power hubs (Li et al., 2022), and e-voting systems (Abegunde, 2022).

To encapsulate, the linchpin of blockchain networks undeniably rests on consensus methodologies. As scholars venture deeper, crafting novel solutions attuned to sectoral nuances, it's evident that blockchain's applications, anchored by these mechanisms, stand poised to instigate industry-wide metamorphoses and catalyze paradigm shifts across disciplines.

Time-Decay Trust Mechanism: A Novel Approach

The concept of a time-decay trust mechanism in blockchain introduces a dynamic trust evaluation system. Traditional trust mechanisms in blockchain are often static, relying on past behaviors to evaluate trustworthiness. However, with time-decay, the trust value of a participant diminishes over time, ensuring that participants remain active and trustworthy in the network. This approach can be particularly beneficial in a reward incentivized HPCC environment for AI, where nodes or participants are rewarded based on their contributions and trustworthiness.

**Method**

Let us contemplate a scenario where we harness the capabilities of high-performance cluster computing to augment the functionality of artificial intelligence. In this context, each participating node's computational contribution to the cluster is a significant factor. The greater the computational power contributed by a node, the higher the probability of accruing rewards. This, essentially, is a manifestation of a PoW (Proof-of-Work) consensus mechanism, where rewards are inherently block production rewards.

With each freshly minted block, we trigger a process of distributed validation. This process primarily follows a weighted random approach, selecting validating nodes based on their computational power. This contributes to a balanced system, where nodes with higher computing capabilities bear a larger share of the validation process. However, this bias towards computational power is carefully moderated over time, employing Bayesian statistical techniques.

The Bayesian approach provides a mechanism for updating the weight assigned to each node based on the accuracy of the blocks they have previously produced. Consequently, this allows for the calculation of a 'trust score', a dynamic metric evaluating each node's reliability. A node's trust score, therefore, evolves over time, with successful and accurate block production increasing the score, while fraudulent or erroneous behavior diminishes it. In essence, the node's computational power and their accuracy in block production feed into their trust score. This score, in turn, influences the likelihood of the node being selected for block production in the future, hence fostering a system of reputation.

While this system might appear to favor computationally powerful nodes, it is important to ensure the inclusion of less powerful units, such as individual personal computers. Their block production chances may seem slim compared to their high-powered counterparts, but the system should still offer a fair opportunity for these units to participate and earn rewards. To implement this, we could consider a statistical 'lottery' system that, despite the correlation between computational power, reputation, and block production chance, allows for a small but significant probability of block production by lower-powered nodes.

The lottery system could be formulated using a negative binomial distribution, where the number of 'failures' before a 'success' is what's being modeled. Here, a 'failure' could be seen as a block production attempt by a high-powered node, and a 'success' as a block production by a lower-powered node. The mean of this distribution would represent the expected number of failures before a success, which could be manipulated to control the rate of block production by lower-powered nodes. By implementing such a framework, we could maintain the integrity of the system while ensuring inclusivity. The detailed mathematical and statistical modeling for this scenario warrants a separate, more exhaustive discussion.

We are essentially looking for a consensus mechanism that primarily favors high computing power, rewards block production based on this power, but also factors in a trust score influenced by a history of correct block validation. Also, we want to give a chance to less powerful computers to participate in the network.

A basic suggestion for such a system:

1. **Block Creation:**
   Each node gets a chance to create a block proportional to its computing power. This is similar to Proof of Work, but without the unnecessary computational waste. We can denote this chance as $C_i$ for node i, which is proportional to its computing power $P_i$.

2. **Block Validation:**
   When a block is created, it's sent to a few randomly chosen nodes for validation. The selection is weighted by the nodes' computing power. This would give more powerful nodes a higher chance to validate blocks, but smaller nodes also get a chance.

3. **Trust Score:**
   A trust score $T_i$ for node i can be maintained, which starts at some default value and is updated based on the node's history of block validations. If a node validates a block correctly (as determined by other nodes), its trust score increases. If a node validates a block incorrectly, its trust score decreases. The Bayesian updating rule could be used in this case.
   Let's say N is the correct validations and M is the incorrect validations for a node. A simple Bayesian trust score could be calculated as:

$$T_i = \frac{N}{N + M}.$$

4. **Block Creation Chance:**
   For each new block, the chance of a node to create it can be calculated as a function of its computing power and its trust score. One possible function could be:

   $$\text{Chance}_i = \alpha C_i + (1 - \alpha) T_i,$$

where $\alpha$ is a parameter between 0 and 1 that determines the relative importance of computing power and trust.

This mechanism allows powerful computers to have an advantage, but it also gives smaller computers a chance to participate, especially if they have a high trust score. The Bayesian trust score helps ensure that nodes are honest and perform their validations correctly.

A simple model based on the previous discussion:

1. **Computing Power and Chance to Create a Block:**
   Let's denote the computing power of node $i$ as $P_i$ and the total computing power of the network as

   $$P_{\text{total}} = P_1 + P_2 + \cdots + P_n$$

for $n$ nodes.
The chance $C_i$ of node $i$ to create a block is then:

$$C_i = \frac{P_i}{P_{\text{total}}}.$$

2. **Trust Score:**
   Each node $i$ also has a trust score $T_i$, which starts at a default value (like 0.5). Whenever node $i$ validates a block correctly, increment a counter $N_i$. Whenever node $i$ validates a block incorrectly, increment a counter $M_i$.
   The trust score $T_i$ after $N_i$ correct validations and $M_i$ incorrect validations can be calculated with the Bayesian rule as:

   $$T_i = \frac{N_i}{N_i + M_i}.$$

To prevent nodes from having a trust score of zero or one (which might cause problems in the calculations), we could add a small number to the numerator and denominator, like:

$$T_i = \frac{N_i + 0.5}{N_i + M_i + 1}.$$

3. **Chance to Create a Block Considering Trust Score:**
   Now, we can calculate the combined chance for a node i to create a block as a weighted average of $C_i$ and $T_i$. One possibility is to use a linear combination:

   $$\text{Chance}_i = \alpha C_i + (1 - \alpha) T_i.$$

Here, $\alpha$ is a parameter between 0 and 1 that we can adjust to give more weight to either computing power or trust.

Note that this is a simplified model, and in practice one would need to handle more complex issues. For example, how to choose which nodes get to validate a block, how to deal with nodes that try to game the system by creating many identities (Sybil attack), and how to handle disagreements about whether a block is valid.

A basic Bayesian trust score update rule is given as:

$$T_i = \frac{N_i + 0.5}{N_i + M_i + 1},$$

where $N_i$ is the number of correct validations and $M_i$ is the number of incorrect validations by node i. This assumes a uniform prior, where every node is equally likely to be trustworthy.

However, Bayesian statistics allow us to use different priors and likelihoods to adjust how the trust score is updated. Few alternatives:

1. **Different Priors:**
   Instead of a uniform prior, we could use a Beta distribution, which is a common choice for Bayesian updating with binary outcomes. The Beta distribution has two parameters, a and b, which we can adjust based on your beliefs about the initial trustworthiness of nodes.
   For example, if we believe that most nodes are trustworthy, we might choose $a > b$. If we believe that most nodes are not trustworthy, we might choose $a < b$.
   The trust score update rule would then be:

$$T_i = \frac{N_i + a}{N_i + M_i + a + b}.$$

2. **Different Likelihoods:**
   Instead of updating the trust score based solely on whether a node validated a block correctly or incorrectly, we could also consider how confidently it made its decision. For example, if a node uses machine learning to validate blocks, it might output a confidence score along with its decision. Nodes that make correct decisions with high confidence could be rewarded more than nodes that make correct decisions with low confidence.

3. **Time-Decaying Trust:**
   In some situations, we might want to give more weight to recent actions than to past actions. One way to achieve this is with a time-decaying factor.
   For instance, we could update the trust score with a rule like:

$$T_i = \gamma \, T_{i,old} + (1 - \gamma) \, T_{i,new},$$

where $T_{i,old}$ is the old trust score, $T_{i,new}$ is the trust score calculated with the basic update rule, and $\gamma$ is a parameter between 0 and 1 that determines the rate of decay.

Using Bayesian methods to calculate trust scores in a consensus protocol is a novel idea and requires significant statistical and computer science understanding. In the case of Markov Chain Monte Carlo (MCMC) methods, these are usually used to estimate complex distributions where direct calculation or sampling is difficult. In particular, the Metropolis-Hastings algorithm is used to create a Markov chain where the stationary distribution of the chain matches the target distribution we're interested in.

If we wanted to calculate trust scores in this way, then we would need to define a model for how nodes validate blocks correctly or incorrectly, and then use the Metropolis-Hastings algorithm to estimate the posterior distribution of the parameters of this model. This is quite a complex task and goes beyond the usual usage of trust scores in blockchain systems.

However, as a very simplified example, let's consider a single node in the network, and let's assume that the chance pof this node validating a block correctly is normally distributed with some unknown mean μand standard deviation σ.

In the Metropolis-Hastings algorithm, we would:

1. Start with some initial guess for the parameters μand σ.

2. Propose a new set of parameters $\mu^{\prime*}$ and $\sigma^{\prime*}$ by adding some random noise to the current parameters.
3. Calculate the likelihood of the data (i.e., the validations made by the node) under the proposed parameters and the current parameters.
4. Calculate the ratio

$$r = \frac{L(\mu^{\prime*}, \sigma^{\prime*})}{L(\mu, \sigma)}.$$

5. If $r \geq 1$, accept the proposed parameters. If $r < 1$, accept the proposed parameters with probability $r$.
6. Repeat steps 2–5 many times to generate a chain of parameter values. After a burn-in period, the distribution of these values should approximate the posterior distribution of the parameters given the data.

This chain could then be used to calculate a Bayesian trust score for the node. For example, the trust score could be the mean $\mu$ of the posterior distribution, which represents the node's expected chance of validating a block correctly.

A detailed mathematical formulation for the Bayesian Trust Score Update using Markov Chain Monte Carlo (MCMC) with a Gaussian distribution and the Metropolis-Hastings algorithm:

1. **Gaussian Distribution Model:**
   - Let $T_i$ denote the trust score of node i, where i represents a specific node in the network.
   - Assume that the trust score $T_i$ follows a Gaussian distribution with mean $\mu_i$ and variance $\sigma_i^2$.
   - Initially, set $\mu_i$ and $\sigma_i^2$ to prior values based on your initial beliefs or assumptions.

2. **Metropolis–Hastings Algorithm:**
   - The Metropolis–Hastings algorithm is used to sample from the posterior distribution of the trust scores based on the correctness of the blocks produced.
   - Let $T_i^{(t)}$ denote the trust score of node i at iteration t of the MCMC algorithm.

### 2.1. Proposal Distribution:

- Define a proposal distribution $Q(T_i^{(t)} \mid T_i^{(t-1)})$ for node i, which generates candidate trust scores given the previous trust score.
- A suitable choice could be a normal distribution centered around the previous trust score with a predefined variance:

$$Q(T_i^{(t)} \mid T_i^{(t-1)}) \sim \mathcal{N}(T_i^{(t-1)}, \text{var}).$$

### 2.2. Acceptance Probability:

- Compute the acceptance probability $\alpha_i$ for the candidate trust score $T_i^{(t)}$ based on the Metropolis–Hastings acceptance criterion:

$$\alpha_i = \min\left(1, \frac{P(T_i^{(t)})}{P(T_i^{(t-1)})} \cdot \frac{Q(T_i^{(t-1)} \mid T_i^{(t)})}{Q(T_i^{(t)} \mid T_i^{(t-1)})}\right).$$

- $P(T_i^{(t)})$ represents the posterior probability of the trust score $T_i^{(t)}$ based on the correctness of the blocks produced by node i.

### 2.3. Update Trust Score:

- Sample a uniform random variable $u \sim \mathcal{U}(0,1)$.
- If $u < \alpha_i$, accept the candidate trust score and set:

$$T_i^{(t)} = T_i^{(t)}.$$

- Otherwise, reject the candidate and keep the previous trust score:

$$T_i^{(t)} = T_i^{(t-1)}.$$

**2.4.** Repeat the above steps for enough iterations to ensure convergence and obtain a set of trust score samples for each node.

3. **Calculation of Bayesian Trust Score:**
   - After the MCMC process, calculate the Bayesian Trust Score for each node based on the collected samples.

- The trust score can be estimated as the mean or median of the obtained trust score samples.

---

## Extending the Proposal With Gibbs Sampling

For simplicity, we keep the assumption that the trust scores follow a Gaussian distribution.

**Gaussian Distribution Model:**
Let $T_i$ denote the trust score of node i, where i represents a specific node in the network. Assume that the trust score $T_i$ follows a Gaussian distribution with mean $\mu_i$ and variance $\sigma_i^2$. Initially, set $\mu_i$ and $\sigma_i^2$ to prior values based on your initial beliefs or assumptions.

**Gibbs Sampling:**
The Gibbs Sampling method is used to sample from the posterior distribution of the trust scores based on the correctness of the blocks produced. This method is effective when the joint distribution is complex but the conditional distributions are easy to sample from.

1. Start with some initial values $T_i^{(0)}$ for the trust scores.
2. For each iteration t of the Gibbs sampling algorithm:

**2.1.** Sample a new value for $\mu_i^{(t)}$ from its conditional distribution given $T_i^{(t-1)}$ and $\sigma_i^{(t-1)}$.

**2.2.** Sample a new value for $\sigma_i^{(t)}$ from its conditional distribution given $T_i^{(t-1)}$ and $\mu_i^{(t)}$.

**2.3.** Sample a new value for $T_i^{(t)}$ from its conditional distribution given $\mu_i^{(t)}$ and $\sigma_i^{(t)}$.

Note: The order in which we update $\mu_i$, $\sigma_i$, and $T_i$ may depend on the specifics of your model and how the conditional distributions are defined.

After the Gibbs sampling process, calculate the Bayesian Trust Score for each node based on the collected samples. The trust score can be estimated as the mean or median of the obtained samples. Make sure to discard a burn-in period and check for convergence.

---

## Introducing Time-Decay Factor

Finally, we can incorporate a time-decay factor into the trust scores. This reflects the idea that more recent behavior is more relevant.

**Gaussian Distribution Model:**
Let $T_i(t)$ denote the trust score of node i at time t. Assume that the trust score $T_i(t)$ follows a

Gaussian distribution with mean $\mu_i(t)$ and variance $\sigma_i^2(t)$. These parameters change over time.

**Time-Decay Model:**

Let $\lambda$ be a decay factor between 0 and 1. The current trust score is a weighted average of the previous trust score and the current behavior:

$$T_i(t) = \lambda\, T_i(t-1) + (1-\lambda)\, B_i(t),$$

where $B_i(t) \in \{0,1\}$ is the new behavior at time t.

This is an exponential moving average.

**Gibbs Sampling:**

Start with initial values $T_i(0)$.

For each iteration (or time step) t:

**2.1.** Update the parameters $\mu_i(t)$ and $\sigma_i^2(t)$ based on the time-decay model.
**2.2.** Sample a new value for $\mu_i(t)$ from its conditional distribution given $T_i(t-1)$, $\sigma_i^2(t-1)$, and $B_i(t)$.
**2.3.** Sample a new value for $\sigma_i^2(t)$ from its conditional distribution given $T_i(t-1)$, $\mu_i(t)$, and $B_i(t)$.
**2.4.** Sample a new value for $T_i(t)$ from its conditional distribution given $\mu_i(t)$ and $\sigma_i^2(t)$.

**Calculation of Bayesian Trust Score:**

After running Gibbs sampling, compute the Bayesian Trust Score as the mean or median of the sampled $T_i(t)$ values.

**Conclusion**

This paper presents a method that merges AI and high-performance computing within a blockchain environment. The idea is straightforward: the more computational power a computer (or node) contributes to the network, the better its chances of earning rewards. These rewards come from creating blocks of data, a core part of how blockchain technology works. It's like a giant, distributed game where the hardest workers get the most points.

But it's not just about brute force. The method also considers a 'trust score'. This score reflects the node's history of creating correct blocks of data. Nodes not only need to be computationally powerful, but they also need to be 'trustworthy', or accurate in their work. In fact, the proposed system even has a way to give less-powerful computers a shot at earning points. This is made possible by a statistical 'lottery' system, which makes sure everyone can participate and win, regardless of power. Like a real-world lottery ticket, it might be a long shot, but the chance is there.

In summary, for this framework to work, we don't need the most high-tech supercomputer; we just have to contribute what we can and be good at what we do. It's a system that rewards hard work and accuracy, but also offers opportunity for all.

The successful integration and implementation of AI and high-performance cluster computing within a blockchain framework, which we have discussed in this paper, is a major milestone that promotes inclusivity, scalability, and sustainability. The potential implications of this model are far-reaching. By providing an equitable and incentivised platform for a wide range of computational power, it bridges the gap between high- and low-powered nodes, fostering a truly democratic, decentralized network.

The introduction of the trust score, which rewards accuracy and consistency, encodes the virtue of trustworthiness into the fabric of the system, ushering in a new level of transparency and reliability. Furthermore, the possibility for smaller, less powerful units to engage and earn rewards by way of a statistical lottery system counteracts potential computational power discrepancies and monopolization.

The environmental footprint of such an approach also bears consideration. By optimizing the usage of existing computational resources, we counteract the need for dedicated, high-powered computing units for each individual task, thus promoting effective resource utilization and potentially reducing e-waste.

Future exploration in modelling, simulations and real-world trials will no doubt shed more light on the strengths and potential challenges of this method. All in all, we believe this framework provides a vital blueprint for creating.

Implications on Sustainable Development

The integration of artificial intelligence (AI) and high-performance cluster computing as described in the paper can support sustainable development in the following ways:

1. Efficient Resource Utilization: Using distributed computation power, AI tasks can be divided and executed across numerous nodes, which could lead to more efficient use of resources and energy. This is particularly significant for training complex AI models, which require substantial computational resources.

2. Scalability: Such a framework allows AI models to leverage additional computational power as the network grows. This scalability is key to handling increased demand and complexity in AI tasks over time.

3. Inclusiveness: By allowing lower-powered nodes to participate, the system utilizes a wide base of available resources, which encourages broader participation and fosters the democratization of AI development.

4. Trust and Validation: The use of a trust score and validation process encourages nodes to operate accurately and honestly. This can improve the reliability and transparency of AI operations on the network.

5. Reducing E-Waste: By utilizing existing computational resources across a distributed network, it reduces the need for every AI project to procure its own dedicated, often very high-powered, hardware. This could contribute to a reduction in electronic waste.

6. Incentivization: Rewarding nodes based on their contribution can incentivize better maintenance and upgrading of computational infrastructure, which could indirectly support the continual advancement of AI capabilities.

It's crucial to consider environmental impacts as well, where sustainable development is not just about effective use of resources, but also minimizing harmful effects like excessive energy consumption, which is a critique often leveled at high-performance computing and blockchain operations.

Appendix:

The consensus algorithm described in the paper involves several complex concepts, including proof-of-work, Bayesian updating, and a lottery system. Below is a simplified Python code snippet that incorporates these ideas:

```python
import random

class Node:
    def __init__(self, id, power):
        self.id = id
        self.power = power
        self.trust_score = 0.5
        self.correct_validations = 0
        self.incorrect_validations = 0

    def update_trust_score(self):
        self.trust_score = (self.correct_validations + 0.5) / (self.correct_validations + self.incorrect_validations + 1)

class Blockchain:
    def __init__(self):
        self.nodes = []
        self.alpha = 0.5

    def add_node(self, node):
        self.nodes.append(node)
```

```python
    def select_block_creator(self):
        chances = [self.alpha * node.power + (1 - self.alpha) * node.trust_score for node in self.nodes]
        total = sum(chances)
        chances = [chance / total for chance in chances]
        return random.choices(self.nodes, weights=chances, k=1)[0]

    def validate_block(self, block_creator):
        for node in self.nodes:
            if node.id != block_creator.id:
                if random.random() < node.power:  # Assume node's power represents its probability of validating a block correctly
                    node.correct_validations += 1
                else:
                    node.incorrect_validations += 1
                node.update_trust_score()

blockchain = Blockchain()
for i in range(10):
    blockchain.add_node(Node(i, random.random()))  # Add 10 nodes with random power

for _ in range(100):
    block_creator = blockchain.select_block_creator()
    blockchain.validate_block(block_creator)
```

This code creates a `Blockchain` class that maintains a list of nodes. Each `Node` has an id, a power (representing its computational power), a trust score, and counters for correct and incorrect block validations.

The `Blockchain` class has a method `select_block_creator` that selects a node to create a block based on a weighted random choice, where the weights are a combination of the node's power and trust score.

The `validate_block` method simulates the validation of a block. Each node (except the one that created the block) has a chance to validate the block correctly based on its power. After each validation, the node's trust score is updated.

Please note that this is a very simplified version of the consensus algorithm described in the paper. It doesn't include the lottery system for less powerful nodes, and the way it simulates block validation is very basic. In a real-world application, the consensus algorithm would be much more complex and would need to handle many more scenarios and edge cases.